\newcommand{\CLASSINPUTbaselinestretch}{1.1}
\pdfoutput=1
\documentclass[10pt, conference, compsocconf]{IEEEtran}
\usepackage[text={7in,9in},centering]{geometry}

\usepackage{epstopdf,epsfig}

\usepackage{caption}

\usepackage{color}
\usepackage[hyphens]{url}
\usepackage[noend]{algorithmic}
\usepackage{paralist}
\usepackage{balance}

\usepackage{xspace}
\newcommand{\etal}{\textit{et al.}\xspace}

\newcommand{\ie}{\textit{i.e.,}\xspace}
\newcommand{\eg}{\textit{e.g.,}\xspace}

\newcommand{\pparagraph}[1]{\medskip \noindent \textbf{#1\ }}
\renewcommand{\paragraph}[1]{\medskip \noindent \textbf{#1.\ }}

\newcommand{\tuple}[1]{\left \langle #1 \right \rangle}

\newcommand{\e}{\ensuremath{e}}  
\newcommand{\g}{\mathbb{G}}

\newcommand{\zz}{\ensuremath{\mathbb{Z}}}

\newcommand{\DLog}{\textrm{DL}}

\newcommand{\pseudonym}{X}
\newcommand{\ts}{\mathrm{ts}}

\newcommand{\iseq}{\ensuremath{\overset{?}{=}}}

\newcommand{\sysname}{\textsc{BackRef}\xspace}

\newcommand{\upon}{\vspace{0pt}\hspace{0ex}\textbf{upon}~}

\newcommand{\pathlength}{\ensuremath{\ell}\xspace}

\newcommand{\FNET}[1]{\ensuremath{\text{{\normalfont $\mathcal {F}_{\textsc{net}^{#1}}$}}}\xspace}

\newcommand{\POR}{\ensuremath{\text{{\normalfont $\Pi_{\textsc{or}}$}}}\xspace}

\newcommand{\FSCS}{\ensuremath{\mathcal F_{\textsc{scs}}}\xspace}
\newcommand{\FREG}{\ensuremath{\mathcal F_{\textsc{reg}}^{\cal N}}\xspace}

\newcommand{\cid}{\ensuremath{\mathit{cid}}\xspace}
\newcommand{\sid}{\ensuremath{\mathit{sid}}\xspace}

\newcommand{\state}{\ensuremath{\mathit{st}}\xspace}

\newcommand{\BConstructOnion}{\WrapOnion}

\newcommand{\UnwrapOnion}{\ensuremath{\mathit{UnwrOn}}\xspace}

\newcommand{\WrapOnion}{\ensuremath{\mathit{WrOn}}\xspace}

\newcommand{\ExtendCircuit}{\ensuremath{\mathit{ExtendCircuit}}\xspace}

\newcommand{\Initiate}{\ensuremath{\mathit{Initiate}}\xspace}
\newcommand{\Respond}{\ensuremath{\mathit{Respond}}\xspace}
\newcommand{\ComputeKey}{\ensuremath{\mathit{ComputeKey}}\xspace}

\newcommand{\sk}{\ensuremath{\mathit{sk}}\xspace}
\newcommand{\pk}{\ensuremath{\mathit{pk}}\xspace}

\newcommand{\nxt}{\ensuremath{\mathit{next}}\xspace}
\newcommand{\prev}{\ensuremath{\mathit{prev}}\xspace}

\newcommand{\getkey}{\ensuremath{\mathit{getkey}}\xspace}

\newcommand{\ready}{\ensuremath{\mathsf{ready}}\xspace}

\newcommand{\register}{\ensuremath{\mathsf{register}}\xspace}
\newcommand{\registered}{\ensuremath{\mathsf{registered}}\xspace}

\newcommand{\setup}{\ensuremath{\mathsf{setup}}\xspace}
\newcommand{\send}{\ensuremath{\mathsf{send}}\xspace}

\newcommand{\respond}{\ensuremath{\mathsf{respond}}\xspace}

\newcommand{\ntor}{\ensuremath{\mathsf{ntor}}\xspace}
\newcommand{\server}{\ensuremath{\mathsf{server}}\xspace}

\newcommand{\relay}{\ensuremath{\mathsf{relay}}\xspace}
\newcommand{\create}{\ensuremath{\mathsf{create}}\xspace}
\newcommand{\created}{\ensuremath{\mathsf{created}}\xspace}
\newcommand{\extend}{\ensuremath{\mathsf{extend}}\xspace}
\newcommand{\extended}{\ensuremath{\mathsf{extended}}\xspace}
\newcommand{\destroy}{\ensuremath{\mathsf{destroy}}\xspace}

\newcommand{\received}{\ensuremath{\mathsf{received}}\xspace}
\newcommand{\data}{\ensuremath{\mathsf{data}}\xspace}
\newcommand{\type}{\ensuremath{\mathsf{type}}\xspace}
\newcommand{\createcircuit}{\ensuremath{\mathsf{createcircuit}}\xspace}
\newcommand{\default}{\ensuremath{\mathsf{default}}\xspace}
\newcommand{\added}{\ensuremath{\mathsf{ex}}\xspace}

\newcommand{\new}{\ensuremath{\mathsf{new}}\xspace}

\newcommand{\ot}{\ensuremath{\overset{\$}{\leftarrow}}\xspace}
\newcommand{\link}[1]{\ensuremath{\overset{#1}{\Longleftrightarrow}}\xspace}

\newcommand{\Circuit}{\ensuremath{\mathcal{C}}\xspace}
\newcommand{\Parties}{\ensuremath{\mathcal{N}}\xspace}

\newcommand{\A}{\ensuremath{\mathcal{A}}\xspace}

\usepackage[cmex10]{amsmath}
\usepackage{amsfonts}

\begin{document}
\title{Introducing Accountability to Anonymity Networks}


\author{\IEEEauthorblockN{Michael Backes\IEEEauthorrefmark{1}\IEEEauthorrefmark{3},
Jeremy Clark\IEEEauthorrefmark{2},
Peter Druschel \IEEEauthorrefmark{3}, 
Aniket Kate\IEEEauthorrefmark{4} and
Milivoj Simeonovski\IEEEauthorrefmark{1}}
\IEEEauthorblockA{\IEEEauthorrefmark{1}Saarland University, Germany\\
\{backes,simeonovski\}@cs.uni-saarland.de}
\IEEEauthorblockA{\IEEEauthorrefmark{4}MMCI, Saarland University, Germany\\
aniket@mmci.uni-saarland.de}
\IEEEauthorblockA{\IEEEauthorrefmark{2}Concordia University, Canada\\
clark@ciise.concordia.ca}
\IEEEauthorblockA{\IEEEauthorrefmark{3}MPI-SWS, Germany\\
druschel@mpi-sws.org}}

\maketitle



%

\begin{abstract}
Many anonymous communication (AC) networks rely on routing traffic through proxy nodes to obfuscate the originator of the traffic. Without an accountability mechanism, exit proxy nodes risk sanctions by law enforcement if users commit illegal actions through the AC network. We present \sysname, a generic mechanism for AC networks that provides practical repudiation for the proxy nodes by tracing back the selected outbound traffic to the predecessor node (but not in the forward direction) through a cryptographically verifiable chain. It also provides an option for full (or partial) traceability back to the entry node or even to the corresponding user when all intermediate nodes are cooperating. Moreover, to maintain a good balance between anonymity and accountability, the protocol incorporates whitelist directories at exit proxy nodes. \sysname offers improved deployability over the related work, and introduces a novel concept of pseudonymous signatures that may be of independent interest.
 
We exemplify the utility of \sysname by integrating it into the onion routing (OR) protocol, and examine its deployability by considering several system-level aspects. We also present the security definitions for the \sysname system (namely, anonymity, backward traceability, no forward traceability, and no false accusation) and conduct a formal security analysis of the OR protocol with \sysname using ProVerif, an automated cryptographic protocol verifier, establishing the aforementioned security properties against a strong adversarial model.
\end{abstract}



\vspace{2ex}
\section{Introduction}
Anonymous communication networks are designed to hide the originator of each message within a larger set of users. In some systems, like DC-Nets~\cite{DC-Net} and Dissent~\cite{Dissent}, the message emerges from aggregating all participants' messages. In other systems, like onion routing~\cite{OnionRouting}, mix networks~\cite{pseudonyms}, and peer-to-peer anonymous communication networks~\cite{ShadowWalker}, messages are routed through volunteer nodes that act as privacy-preserving proxies for the users' messages. We call this latter class proxy-based anonymous communication (AC) networks and concentrate on it henceforth.

Proxy-based AC networks provide a powerful service to their users, and correspondingly they have been the most successful AC networks so far~\cite{Tor,mixmaster}.
However the nature of the properties of the technology can sometimes be harmful for the nodes serving as proxies. If a network user's online communication results in a criminal investigation or a cause of action, the last entity to forward the traffic may become embroiled in the proceedings~\cite{GermanBlogger,Accused,BloggerPresentation}, whether as the suspect/defendant or as a third party with evidence. While repudiation in the form of a partial or full traceability has never been a component of any widely-deployed AC network, it may become the case that new anonymity networks, or a changing political climate, initiate an interest in providing a verifiable trace to users who misuse anonymity networks according to laws or terms of service.

While several proposals~\cite{RevocableAnonimity,SelectTraceable,diaz2007accountable,GolleRMN,ExitNodeRep,Nymble,Blacklisting} 
have been made to tackle or at least to mitigate this problem under the umbrella term of {\em accountable anonymity},
as we discuss in the next section some of them are broken, 
while others are not scalable enough for deploying in low latency AC networks.

\paragraph{Contributions}
In this work, we design \sysname, a novel practical repudiation mechanism 
for anonymous communication, which has advantages in terms of deployability and efficiency over the literature.
To assist in the design of \sysname, we propose a concept of pseudonymous signatures, which employ pseudonyms (or half Diffie-Hellman exponents) as temporary public keys (and corresponding temporary secrets) employed or employable in almost all AC networks for signing messages. These pseudonym signatures are used to create a verifiable {\em pseudonym-linkability} mechanism where any proxy node within the route or path, {\em when required}, can verifiably reveal its predecessor in time-bound manner. We use this property to design a novel repudiation mechanism, which allows each proxy node, in cooperation with the network, to issue a cryptographic guarantee that a selected traffic flow can be traced back to its originator (\ie predecessor node) while maintaining the eventual forward secrecy of the system. 

Unlike the related work, which largely relies on group signatures and/or anonymous credentials, \sysname avoids the logistical difficulties of organizing users into groups and arranging a shared group key, and does not require access to a trusted party to issue credentials.
While \sysname is applicable to all proxy-based AC networks, we illustrate its utility by applying it to the onion routing (OR) protocol. We observe that it introduces a small computational overhead and does not affect the performance of the underlying OR protocol.
\sysname also includes a {\em whitelisting} option; \ie if a exit node considers traceability  to one or more web-services unnecessary, 
then it can include those services in a {\em whitelist} directory such that accesses to those are not logged.

We formally define the important properties of the \sysname network.
In particular, we formalize anonymity and no forward traceability as observational equivalence relations,
and backward traceability and no false accusation as trace properties.
We conduct a formal security analysis of \sysname using ProVerif, an automated cryptographic protocol verifier, establishing the aforementioned security and privacy properties against a strong adversarial model. 
We believe both the definitions and the security analysis are of independent interest, since they are the first for the OR protocol.

\paragraph{Organization} 
In Section~\ref{sec:Background}, we discuss the anonymous communication networks, 
and consider the related work. In Section~\ref{sec:Overview}, 
we describe our threat model and system goals, and present our key idea,
while in Section~\ref{sec:Traceability}, we incorporate the \sysname mechanism in the OR protocol.
We discuss important systems issues in Section~\ref{sec:TSysAspects}, and 
we briefly analyze the security and privacy properties of the \sysname mechanism in Section~\ref{sec:Analysis}.
\section{Background and Related Work}\label{sec:Background}
Anonymous communication (AC) networks aim at protecting personally identifiable information (PII),
in particular the network addresses of the communicating parties
by hiding correlation between input and output messages at one or more network entities. 
For this purpose, the AC protocols employ techniques
such as using a series of intermediate routers and layered encryptions to obfuscate the source of a communication, 
and adding fake traffic to make the `real' communication difficult to extract.

\paragraph{Anonymous Communication Protocols}
Single-hop proxy servers, which relay traffic flows, enable a simple form of anonymous communication. However anonymity in this case requires, at a minimum, that the proxy is trustworthy and not compromised, and this approach does not protect the anonymity of senders if the adversary inspects traffic through the proxy~\cite{GWB}. Even with the use of encryption between the sender and proxy server, timing attacks can be used to correlate flows.

Starting with Chaum~\cite{pseudonyms}, several AC technologies have been developed in the last thirty years to provide stronger anonymity not dependent on a single entity~\cite{Tor,OnionRouting,mixmaster,Crowds,Dissent,DC-Net,FreedomNetwork,Kate07pairing-basedonionimproved,Sphinx,JAPorANON,Mixminion,Tarzen}. Among these, mix networks~\cite{pseudonyms,mixmaster} and onion routing~\cite{Tor} have arguably been most successful. Both offer user anonymity, relationship anonymity and unlinkability~\cite{anon_terminology}, but they obtain these properties through differing assumptions and techniques. 

An onion routing (OR) infrastructure involves a set of {\em routers} (or {\em OR nodes}) that relay traffic, a \textit{directory service} providing status information for OR nodes, and \textit{users}. Users benefit from anonymous access by constructing a \textit{circuit}---a small ordered subset of OR nodes---and routing traffic through it sequentially. The crucial property for anonymity is that an OR node within the built circuit is not able to identify any portion of the circuit other than its predecessor and successor. The user sends messages (to the first OR node in the circuit) in a form of an \textit{onion}---a data structure multiply encrypted by symmetric session keys (one encryption layer per node in the circuit). The symmetric keys are negotiated during an initial \textit{circuit construction} phase. This is followed by a second phase of  {\em low latency} communication (opening and closing streams) through the constructed circuit for the session duration. An OR network does not aim at providing anonymity and unlinkability against a global passive observer, which in theory can  analyze end-to-end traffic flow. Instead, it assumes an adversary that adaptively compromises a small fraction of OR nodes and controls a small fraction of the network. 

A mix network achieves anonymity by relaying messages through a path of mix nodes. The user encrypts a message to be partially decrypted by each mix along the path. Mix nodes accept a batch of encrypted messages, which are partially decrypted, randomly reordered, and forwarded. Unlike onion routing, an observer is unable to link incoming and outgoing messages at the mix node; thus, mix networks provide anonymity against a powerful global passive adversary. In fact, as long as a single mix node in the user's path remains uncompromised, the message will maintain some anonymity. However, batching of messages at a mix node introduces inherent delays, making mix networks unsuitable for low-latency, interactive applications (\eg web browsing, instant messaging). When used, it is for latency-tolerant applications like anonymous email.

\subsection{Accountable Anonymity Mechanisms}\label{sec:RelatedWork}

The literature has examined several approaches for adding accountability to AC technologies, allowing: misbehaving users to be selectively traced~\cite{RevocableAnonimity,SelectTraceable,diaz2007accountable}, exit nodes to deny originating traffic it forwards~\cite{GolleRMN,ExitNodeRep}, misbehaving users to be banned~\cite{Nymble,Blacklisting}, and misbehaving participants to be discovered~\cite{Dissent,Dissent++,Verdict}. All of these approaches either require users to obtain credentials or do not extend to interactive, low-latency, internet-scale AC networks. A number also partition users into subgroups, which reduces anonymity and requires a group manager. \sysname does not require credentials, subgroups, and is compatible with low-latency AC networks like onion routing, adding minimal overhead.

Kopsell \etal~\cite{RevocableAnonimity} propose traceability through threshold group signatures. A user logs into the system to join a group, signs messages with a group signature, and a group manager is empowered to revoke anonymity. The system also introduces an external proxy to inspect all outbound traffic for correct signatures and protocol compliance. The inspector has been criticized for centralizing traffic flows, which enables DOS, censorship, and increases observability~\cite{BypassRevocationSchemes}.

Von Ahn \etal~\cite{SelectTraceable} also use group signatures as the basis for a general transformation for traceability in AC networks and illustrate it with DC networks. Users are required to register as members of a group capable of sending messages through the network. Our solution can be viewed as a follow-up to this paper, with a concentration on deployability: we do not require users to be organized into groups or introduce new entities, and we concentrate on onion routing.

Diaz and Preneel~\cite{diaz2007accountable} propose traceability through issuing anonymous credentials to users and utilizing a traitor tracing scheme to revoke anonymity. It is tailored to high-latency mix networks and requires a trusted authority to issue credentials---both impede deployability. Danezis and Sassaman~\cite{BypassRevocationSchemes} demonstrate a bypass attack on this and the Kopsell \etal scheme~\cite{RevocableAnonimity}. The attack is based on the protocols' assumption that there can be no leakage of information from inside the channel to the world unless it passes through the verification step. This attack is only applicable for the family of protocols where traceability property is ensured. In our protocol we do not claim ensured traceability therefore this attack is out of the scope of \sysname.


Short of revoking to the anonymity of misbehaving users, techniques have been proposed to at least allow exit nodes to deny originating the traffic. Golle~\cite{GolleRMN} and Clark \etal~\cite{ExitNodeRep} pursue this goal, with the former being specific to high-latency mix networks and the latter requiring anonymous credentials. Tor offers a service called ExoneraTor that provides a record of which nodes were online at a given time, but it does not explicitly prove that a given traffic flow originated from Tor. Other techniques, such as Nymble~\cite{Nymble} and its successors (see a survey~\cite{Blacklisting}), enable users to be banned. However these systems inherently require some form of credential or pseudonym infrastructure for the users, and also mandate web-servers to verify user requests. Finally, Dissent~\cite{Dissent} and its successors~\cite{Dissent++,Verdict} presents an interesting approach for accountable anonymous communication for DC Nets~\cite{DC-Net}, however even when highly optimized~\cite{Dissent++}, DC Nets are not competitive for internet-scale application. 


\begin{figure*}[t!]
\centering
\includegraphics[trim=2cm 7cm 0cm 7cm,clip=true,angle=0,scale=0.55]{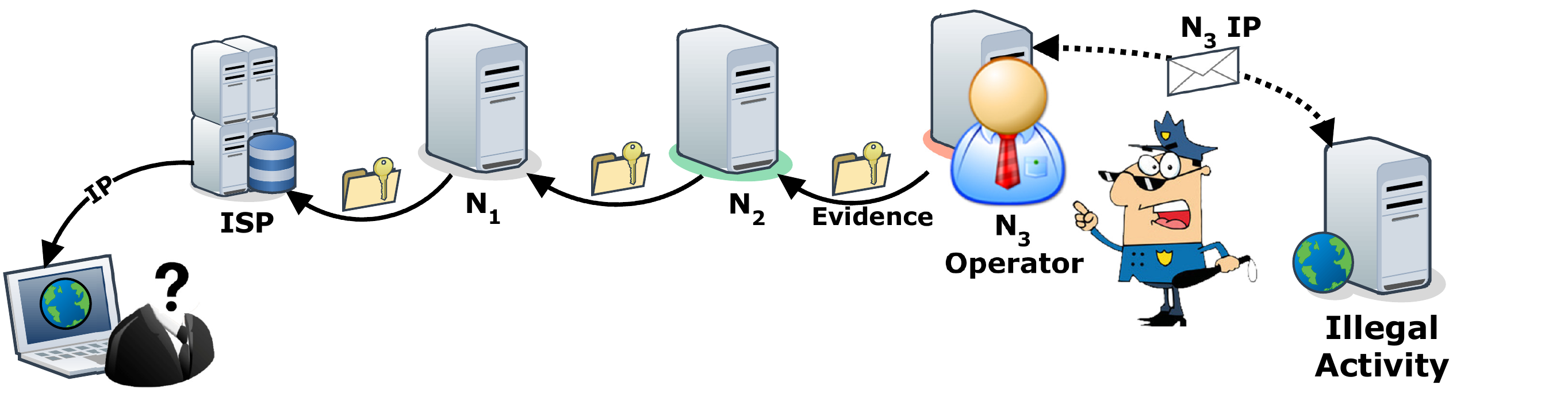}
\caption{Backward Traceability Verification}
\label{BackwardTraceability}

\end{figure*}

\section{Design Overview}\label{sec:Overview}
In this section we describe our threat model and system goals, 
and present our key idea and design rationale.

\subsection{Threat Model and System Goals}\label{sec:Goals}
We consider the same threat model as the underlying AC protocol in which we wish to incorporate the \sysname\ mechanism. 
Our active adversary $\A$ aims at breaking some anonymity property 
by determining the ultimate source and/or destination of a communication stream
or breaking unlinkability by linking two communication streams of the same user.
We assume that some, but not all, of the nodes in the path of the communication stream are compromised by the adversary $\A$,
who knows all their secret values, and is able to fully control their functionalities. 
For high latency AC networks like mix networks, we assume that the adversary can also observe all traffic in the network, 
as well as intercept and inject arbitrary messages,
while for low latency AC networks like onion routing, we assume the adversary can observe, intercept, and inject traffic
in some parts of the network.

While maintaining the anonymity and unlinkability properties of the AC network, 
we wish to achieve the following goals when incorporating \sysname in an AC network:

\begin{LaTeXdescription}

\item [Repudiation:]
For a communication stream flowing through a node, 
the node operator should be able to prove that the stream
is coming from another predecessor node or user.

\item [Backward traceability:]
Starting from an exit node of a path (or circuit), it should be possible to trace the source of a communication stream
when all nodes in the path verifiably reveal their predecessors.

\begin{sloppypar}
\item[No forward traceability:]
For a  compromised node, 
it should not be possible for the adversary $\A$ to use \sysname to verifiably trace its successor 
in any completed anonymous communication session through it.
\end{sloppypar}

\item [No false accusation:]
It should not be possible for a compromised node to corrupt the \sysname\ mechanism  to trace a communication stream:
\begin{compactenum}
\item to a path different from the path employed for the stream, and 
\item to a node other than its predecessor in the path.
\end{compactenum}

\end{LaTeXdescription}

\paragraph{Non-Goals}
We expect our accountability notion to be reactive in nature.
We do not aim at proactive accountability and do not try to stop an illegal activity in an AC network 
in a proactive manner, as we believe perfect white- or black-listing of web urls and content to be an infeasible task. 
Moreover, some nodes may choose not to follow the \sysname mechanism locally 
(e.g., they may not maintain or share the required evidence logs),
and backward traceability to the user cannot be ensured in those situations;
nevertheless, the cooperating nodes can still prove their innocence in a verifiable manner.

Due to its reactive nature, our repudiation mechanism inherently requires evidence logs containing verifiable routing information.
Encrypting these logs and regularly rotating the corresponding keys can provide us eventual forward secrecy~\cite{LaPa_07}.
However, we cannot aim for {\em immediate} forward secrecy 
due to the inherently eventual forward secret nature of the encryption mechanism.

\subsection{Design Rationale and Key idea}\label{sec:Pseudonyms}
Fig.\ \ref{BackwardTraceability} presents a general expected architecture
to achieve the above mentioned goals.
It is clear the network level logs as well as the currently cryptographic mechanism in the AC networks
cannot be used for verifiably backward traceability purpose 
as they cannot stop false accusations (or traceability) by compromised nodes:
a compromised node can tamper with its logs to intermix two different paths
as there is no cryptographic association between different parts of an AC path.

We observe that almost all OR circuit construction protocols~\cite{
Kate07pairing-basedonionimproved,KateSphinx,Ace,Certificateless,ntor,LaPa_07} 
(except TAP)  and  
mix network protocols~\cite{CaLy_05,Sphinx,Drac,mixmaster,Mixminion,FixedMinx} 
employ (or can employ\footnote{Although some these have been defined using RSA encryptions, 
as discussed in \cite{Sphinx} they can be modified to work in the discrete logarithm (\DLog) setting.}) 
an element of a cyclic group of prime order
satisfying some (version of) Diffie-Hellman assumption 
as an authentication challenges or randomization element
per node in the path.
In particular, it can be represented as
$\pseudonym = g^{x}$,
where $g$ is a generator of a cyclic group $\g$ of prime order $p$ with the security parameter $\kappa$ 
and 
$x \in_R \zz_p$ is a random secret value known only to the user. 
This element is used by each node on the path to derive
a secret that is shared with the user
and is used to extract a set of (session) keys for encryption and integrity protection.
In the literature, these authentication challenges $\pseudonym$ are known as user {\em pseudonyms}.

The key idea of our \sysname\ mechanism is to use 
these pseudonyms $\pseudonym = g^x$ and the corresponding secret keys $x$ 
as signing key pairs to sign pseudonym's for successor nodes at entry and middle nodes,
and to sign the communication stream headers at the exit nodes.
Signatures that use $(x,g^x)$ as the signing key pair are referred to as {\em pseudonym signatures}. 
As pseudonyms are generated independently for every single node, and 
the corresponding secret exponents are random elements of $\zz_p$, 
they do not reveal the user's identity.
Moreover, it also is not possible to link two or more pseudonyms to a single identity.
Therefore,  
pseudonym signatures become particularly useful in our \sysname mechanism,
where users utilize them to sign messages without being identified by the verifier.

We can employ a CMA-secure~\cite{CMA} signature scheme against a computationally bounded adversary 
(with the security parameter $\kappa$) such that, 
along with the usual existential unforgeability, the resultant pseudonym signature scheme 
satisfies the following property:
\begin{LaTeXdescription}
\item [Unconditional signer anonymity:] The adversary cannot determine a signer's identity, 
even if it is allowed to obtain signatures on an unbounded number of messages of its choice.
\end{LaTeXdescription}


We use such temporary signing key pairs (or pseudonym signatures) to sign consecutively employed pseudonyms in an AC path
and the web communication requests leaving the AC path.
Pseudonym signatures provide linkability between the employed pseudonyms 
and the communicated message on an AC path.
However, these pseudonyms are not sufficient to link the node employed in the AC path: 
for a pseudonym received by a node, its predecessor node can always deny sending the pseudonym in the first place. 
We solve this problem by introducing {\em endorsement signatures}: 
We assume that 
every node signs the pseudonym while sending it to the successor
so that it cannot plausibly deny this transfer during backward tracing.

\subsection{Scope of Solution}

To understand the scope of \sysname, first consider traceability in the context of the simplest AC network: a single-hop proxy. Any traceability mechanism from the literature implicitly assumes a solution to the problem of how users can be traced through a simple proxy. We dub this the `last mile' problem. The proxy can keep logs, but this requires a trusted proxy. Alternatively the ISP could observe and log relevant details about traffic to the proxy, requiring trust in the ISP. The solution more typically used in the literature is to assume individual users have digital credentials or signing keys---essentially some form of PKI is in place to certify the keys of individual users.~\cite{RevocableAnonimity,SelectTraceable,diaz2007accountable,GolleRMN,ExitNodeRep}

None of these last mile solutions are particularly attractive. The assumption of a PKI provides the best distribution of trust but short-term deployment appears infeasible. We believe the involvement of ISPs is the most readily deployable. Such a solution involves an ISP with a packet attestation mechanism~\cite{packetsISP} which acts as a trusted party capable of proving the existence of a particular communication. We discuss the packet attestation mechanism further in Section~\ref{sec:TSysAspects}.

For selected traffic flows, \sysname provides traceability to the entrance node. This is effectively equivalent to reducing the strong anonymity of a distributed cryptographic AC network to the weak anonymity of a single hop proxy. For full traceability, we then must address the 'last mile' problem: tracing the flow back to the individual sender. Thus \sysname is not a full traceability mechanism, but rather an essential component that can be composed with any solution to the last mile problem. While we later discuss a solution that involves ISPs, we emphasize that \sysname itself is concentrated on, arguably, the more difficulty problem of offering ensured traceability within the AC network.

\section{Repudiation (or Traceability)}\label{sec:Traceability}
In this section, we present our \sysname repudiation scheme.
For ease of exposition, we include our scheme in an OR protocol instead of
including it in the generic AC protocol. Nevertheless, our scheme is applicable to almost all
AC protocols mentioned in Section~\ref{sec:Pseudonyms}.
We start our discussion with a brief overview of the OR protocol in the Tor notions~\cite{tor-spec}.
We then discuss the protocol flow for \sysname, describe our cryptographic
components, and present a formal pseudocode.

\subsection{The OR Protocol: Overview}\label{sec:OROverview}
The OR protocol is  defined in two phases: circuit construction and streams relay.

\paragraph{OR Circuit Construction} 
The circuit construction phase involves the user onion proxy (OP) randomly selecting a short circuit of (\eg 3) OR nodes, 
and negotiating a session key with each selected OR node using one-way authenticated key exchange (1W-AKE)~\cite{ntor}
such as the \ntor protocol. (We refer the readers to Appendix~\ref{sec:ntor} for more details.)
When a user wants to \create a circuit with an OR node $N_1$, she runs the \Initiate procedure of the \ntor protocol to generate 
 and send an authentication challenge to $N_1$. Node $N_1$ then runs the \respond procedure and returns the authentication response. 
Finally, the user uses the \ComputeKey procedure of \ntor along with the response to authenticate $N_1$ and to compute a session key with it. To \extend the circuit further, the user sends an \extend request to $N_1$ specifying the address of the next node $N_2$ and a new \ntor authentication challenge for $N_2$.  The process continues to until the user exchanges the key with the exit node $N_3$.

\begin{sloppypar}
\paragraph{Relaying Streams} 
Once a circuit (denoted as $\tuple{U \leftrightarrow N_1\leftrightarrow N_2 \leftrightarrow N_3}$) has been constructed
through $N_1$, $N_2$ and $N_3$, the user-client $U$ routes traffic through the circuit
using onion-wrapping \WrapOnion and onion-unwrapping \UnwrapOnion procedures.
\WrapOnion creates a layered encryption of a payload (plaintext or onion) given an ordered list of (three) session keys. 
\UnwrapOnion removes one or more layers of encryptions from an onion to output a
plaintext or an onion given an input onion and a ordered list of one or more session keys. 
To reduce latency, many of the user's communication streams employ the same circuit~\cite{Tor}.
\end{sloppypar}

The structure and components of communication streams may vary with the network protocol. For ease of exposition, 
we assume the OR network uses TCP-based communication in the same way as Tor, 
but our schemes can easily be adapted for other types of communication streams.

In Tor, the communication between the user's TCP-based application and her Tor proxy takes place via SOCKS. 
To open a communication stream (i.e., to start a TCP connection to some web server and port), 
the user proxy sends a \textit{relay begin} cell (or packet) over the circuit to the exit node $N_3$. 
When $N_3$ receives the TCP request, it makes a standard TCP handshake with the web server. 
Once the connection is established, $N_3$ responds to the user with a \textit{relay connected} cell.
The user then forwards all TCP stream requests for the server as \textit{relay data} cells to the circuit. 
(See \cite{Tor,tor-spec} for a detailed explanation.) 
\subsection{The BackRef Protocol Flow} \label{sec:TOverview}
Consider a user $U$ who  wishes to construct an OR circuit $\tuple{U \leftrightarrow N_1 \leftrightarrow N_2 \leftrightarrow N_3}$,
and use it to send communication stream $m$.
\sysname adds the repudiation mechanism as a layer on the top of the existing OR protocol.
We assume that every OR node possesses a signing (private) key for which the corresponding verification (public) key
is publicly available through the OR directory service.

The corresponding OR protocol with the \sysname scheme works according to the following five steps:

\pparagraph{1.~Circuit construction with an entry node:}
The user $U$ creates a circuit with the entry node $N_1$ using the \ntor protocol.
If the user is an OR node,
then it endorses its pseudonym $\pseudonym_1$ by signing it with its public key
and sending the signature along with $\pseudonym_1$.
 
However, if the user $U$ is not an OR node, it cannot endorse the pseudonym $\pseudonym_1$
as no public-key infrastructure (PKI) or credential system is available to him.
We solve this endorsement problem by entrusting the ISP with a packet attestation mechanism~\cite{packetsISP} 
such that the ISP can prove that a pseudonym was sent by $U$ to $N_1$.
We discuss the packet attestation mechanism in Section~\ref{sec:TSysAspects}.

%


\pparagraph{2.~Circuit extension:}
To \extend a circuit to $N_2$, $U$ generates a new pseudonym $\pseudonym_2$ of an \ntor instance,
signs $\pseudonym_2$ and the current timestamp with the secret value $x_1$ associated with $\pseudonym_1$, 
and sends an extend request to $N_1$ along with the identifier for $N_2$, $\{\pseudonym_2||\ts_{x_2}\}_{\sigma_{\pseudonym_1}}$ and a timestamp $\ts_{x_2}$. 
Notice that the extension request is encrypted by a symmetric session key negotiated between $U$ and $N_1$. 

\begin{sloppypar}
Upon receiving a message, $N_1$ decrypts and verifies $\{\pseudonym_2||\ts_{x_2}\}_{\sigma_{\pseudonym_1}}$ using the previously received
 pseudonym $\pseudonym_1$ and timestamp. We call this verification {\em pseudonyms linkability verification}.
If the signature is valid, it creates an evidence record as discussed in Step 4, 
signs $\pseudonym_2$ using its private key to generate $\{\pseudonym_2||\ts_2\}_{\sigma_{\sk_2}}$
and sends a circuit \create request to the node $N_2$ with  $\{\pseudonym_2||\ts_2\}_{\sigma_{\sk_2}}$.
\end{sloppypar}

Node $N_2$, upon receiving a circuit creation request along with $\{\pseudonym_2||ts_2\}_{\sigma_{\sk_2}}$,
verifies the signature. Upon a successful verification, 
it replies to $N_1$ with an \ntor authentication response for the OR key agreement 
and generates the OR session key for its session with (unknown) user $U$. 
$N_1$ sends the authentication response back to $U$ using their OR session,
who then computes the session key with $N_2$ and continues to build its circuit
to $N_3$ in a similar fashion.

Notice that we carefully avoid any conceptual modification of the OR circuit construction protocol;
the above signature generation and verification steps are the only adjustments that \sysname\ makes to
this protocol. 

\pparagraph{3.~Stream verification:} Once a circuit $\tuple{U \leftrightarrow N_1 \leftrightarrow N_2 \leftrightarrow N_3}$ 
has been established, the user $U$ can utilize it to send her web stream requests. To open a TCP connection, 
the user sends a {\it relay begin} cell to the exit node $N_{3}$ through the circuit.
The user $U$ includes a pseudonym signature (or stream request signature) 
on the cell contents signed with the secret exponent $x_{3}$ of $\pseudonym_{3}$.
The user also includes a timestamp in her stream request.
 When the {\it relay} cell reaches the exit node $N_{3}$, 
the exit node verifies the pseudonym signature with $\pseudonym_{3}$.
Once the verification is successful and the timestamp is current, $N_{3}$ creates the evidence log (Step $4$) 
and proceeds with the TCP handshake to the destination server. 
The {\it relay} stream request is discarded otherwise
This stream verification helps $N_3$ to prove linkability between its handshakes with the destination server 
and the pseudonym $\pseudonym_{3}$ it received from $N_2$. 

When a whitelist directory exists, the exit node first consults the directory and if the request ($\ie$ web stream request) is whitelisted, the exit node just forwards it to the destination server. In such a case, the exit node does not require any signature verification and also does not create an evidence log. 
We further discuss the server whitelisting in Section~\ref{sec:Whitelist}.

\pparagraph{4.~Log generation:}
After every successful pseudonym linkability or stream verification, 
the evidence record is created. 
A pseudonym linkability verification evidence record associates linkability between two pseudonyms $\pseudonym_i$ and $\pseudonym_{i+1}$
and an endorsement signature on $\pseudonym_i$,
while a stream verification evidence record associates a stream verification with an endorsement signature on $\pseudonym_3$ 
for $N_3$.


\pparagraph{5.~Repudiation or traceability:}
The verifier contacts the exit node $N_3$ with the request information (e.g., IP address, port number, and timestamp) for a malicious stream coming out of the exit node $N_3$.
The operator of $N_3$ can determine a record using the stream request information.
This evidence record verifiably reveals the identity of the middle node $N_2$.\\
As an optional next step, using the evidence records, 
it is possible for $N_2$ to verifiably reveal the identity of the predecessor node $N_1$.
Then, the last mile of a full traceability is to reach from $N_1$ to the user $U$ in a verifiable manner using the evidence record on $N_1$ and the request information on the  ISP~\cite{packetsISP}. 
When the user $U$ is an OR node a record at $N_1$ is sufficient and the last mile problem does not exist.

\subsection{Cryptographic Details} \label{sec:TraceabilityDetails}
For pseudonym and endorsement signatures, we use the short signature scheme of Boneh, Lynn and Shacham (BLS) \cite{BLSsig}. We recall the BLS signature scheme in Appendix~\ref{App:BLS}.
%
%
We choose the BLS signature scheme due to the shorter size of their signatures;
however, if signing and verification efficiency is more important, 
we can choose faster signature schemes such as \cite{fastSignature}.



\paragraph{Circuit Extension}
To extend the circuit $\tuple{U \leftrightarrow N_1 }$ to the next hop $N_2$, 
the user $U$ chooses $x_2 \in_R \zz_p$ and generates a pseudonym $\pseudonym_2 = g_2^{x_2}$,
where $g_2 \in \g_2$.
$U$ then signs the pseudonym $\pseudonym_2$ with pseudonym $\pseudonym_1$ as public key. Also we include the current timestamp value $\ts_{x_2}$ in the signature
$\sigma_{\pseudonym_1} = H(\pseudonym_2||\ts_{x_2})^{x_1}$
Upon receiving the signed pseudonym $\left\{\pseudonym_2||\ts_{x_2}\right\}_{\sigma_{\pseudonym_1}}$along with the timestamp $\ts_{x_2}$, the node $N_1$ checks if the timestamp is current and verifies it as follows:
\begin{equation*}
e(H(\pseudonym_2||\ts_{x_2}),\pseudonym_1) \iseq e(\sigma_{\pseudonym_{1}}, g_2)
\end{equation*}


\paragraph{Pseudonym endorsement} 
After successful verification, $N_1$ creates an endorsement signature $\sigma_{{1}} = H(\pseudonym_{2}||\ts_2)^{\sk_{1}}$ 
for pseudonym $\pseudonym_2$ and current timestamp $\ts_2$ using its signing key $\sk_1$ and sends it along with $\pseudonym_2$ and $\ts_2$ to $N_2$.

The node $N_2$ then follows the pseudonym endorsement step. 
%
Upon receiving the signed pseudonym $\{\pseudonym_{2}||\ts_2\}_{\sigma_{1}}$, 
the exit node $N_{2}$ verifies it as follows:
\begin{equation*}
e(H(\pseudonym_2||\ts_2),\pk_1) \iseq e(\sigma_1, g_2).
\end{equation*}
On a successful verification, $N_{2}$ continues with the OR protocol.

\paragraph{Stream verification} 
To generate a stream request signature, the user signs the stream request 
(\ie selected contents of the {\it relay begin} cell) using the pseudonym $\pseudonym_{3}=g_{2}^{x_{3}}$ where 
$x_{3}$ is the secret corresponding to $\pseudonym_{3}$. 
For contents of the \textit{relay} cell $m$ = $\{\mathrm{address}\| \mathrm{port}\| \ts_{x_m}\}$,
the stream request signature $\sigma_{X_{3}}$ is defined as
\begin{equation*}
\sigma_{\pseudonym_{3}} = H(m)^{x_3}.
\end{equation*}
The user sends the signature along with the {\it relay} cell and the current timestamp $\ts_{x_m}$ to the exit node 
through the already-built circuit. 

Once the signed stream request reaches $N_{3}$, 
it verifies the signature as follows:
\begin{equation}\label{eq:}
e(H(m),\pseudonym_3) \iseq e(\sigma_{\pseudonym_{3}},g_2).
\end{equation}

Upon a successful verification, the exit node $N_{3}$ proceeds with the TCP handshake.
A verified request allows the node to link $\pseudonym_3$ 
and the request. 

\paragraph{Log generation}
After every successful pseudonym or stream verification, an evidence record is added to the evidence log. 
The evidence records differ with nodes' positions within a circuit, and we define two types of evidence logs.

\begin{LaTeXdescription}

\begin{sloppypar}
\item [{\it Exit node log}:] 
For every successful stream verification, an evidence record is added to the evidence log at the exit node. 
%
A single evidence record consists of
the signature on $\pseudonym_{3}$ (i.e., $\{\pseudonym_{3}||\ts_3\}_{\sigma_2}$), and
the stream request ($m$ = $\{\mathrm{address}\| \mathrm{port}\| \ts_{x_m}\}$) coupled by the pseudonym signature
$\{m\}_{\sigma_{\pseudonym_{3}}}$ and the timestamp $\ts_{x_m}$. \\

\end{sloppypar}



\item [{\it Middle and entry node log}:]
The middle and entry node evidence record 
comprises two pseudonyms $\pseudonym_i$, $\pseudonym_{i+1}$, and a timestamp value $\ts_{x_{i+1}}$ coupled with the appropriate signatures
and the IP address of $N_{i-1}$. The pseudonym $\pseudonym_i$ is coupled with an endorsement signature $\left\{\pseudonym_i||\ts_i\right\}_{\sigma_{i-1}}$ from node $N_{i-1}$, and the pseudonym $\pseudonym_{i+1}$ is coupled by a pseudonym signature $\left\{\pseudonym_{i+1}||\ts_{x+1}\right\}_{\sigma_{\pseudonym_i}}$. 



When the user is not an OR node and does not posse a verifiable signature key pair,
the corresponding record at $N_1$ consists of a signed pseudonym $\left\{\pseudonym_2||\ts_{x_2}\right\}_{\sigma_{\pseudonym_1}}$, pseudonym $\pseudonym_1$, timestamp value $\ts_{x_2}$, and the IP of the user.


\end{LaTeXdescription}

\paragraph{Repudiation or traceability}
Given the server logs of a stream request, an evidence record corresponding to the stream request can be obtained.
In the first step, it is checked whether the timestamp matches the stream request under observation.
In the next step, the association between the stream request and the pseudonym of the exit node $\pseudonym_3$
is verified using the pseudonym signature. 
Then, the association of the pseudonym $\pseudonym_3$ and $N_2$ is 
checked using the pseudonym endorsement signature. 

Given the pseudonym $\pseudonym_3$ and a timestamp $\ts_{x_m}$, the backward traceability verification at node $N_2$ is carried out as follows:
\begin{enumerate}
\item Do a lookup in the evidence log to locate the signed pseudonym $\left\{\pseudonym_3||\ts_{x_3}\right\}_{\sigma_{\pseudonym_2}}$ and the timestamp $\ts_{x_3}$,  
where $\pseudonym_3$ is the lookup index. 

\item Compare the timestamps ($\ts_{x_m}$ and $\ts_{x_3}$) under observation and prove the linkability between $\pseudonym_2$ and $\pseudonym_3$ by verifying the signature $\left\{\pseudonym_3||\ts_{x_3}\right\}_{\sigma_{\pseudonym_2}}$.

\item If verification succeeds, reveal the IP address of the node $N_1$ who has forwarded $\pseudonym_2$ 
and verify  $\left\{\pseudonym_2||\ts_2\right\}_{\sigma_{1}}$ with $\pk_1$.
\end{enumerate}

The above three steps can be used repeatedly to reach the entry node.
However, they cannot be used to verifiably reach the user if we do not assume any public key and credential infrastructure
for the users. 
Instead, our protocol relies on the ISP between user $U$ and $N_1$ 
to use packet attestation~\cite{packetsISP} to prove that the pseudonym $X_1$ was sent from $U$ to $N_1$. 

\subsection{Exit Node Whitelisting Policies}\label{sec:Whitelist}
To provide a good balance between anonymity and accountability, 
we include a whitelisting option for exit nodes. 
This option allows a user to avoid the complete verification and logging mechanisms
if her destination is in the whitelist directory of her exit node. In particular, we categorize the destinations into two groups:

\pparagraph{Whitelisted destinations:}
For several destinations such as educational {\tt .edu} websites, an exit node may find traceability to be unnecessary.
The exit node includes such destinations in a whitelist directory 
such that, for these destinations, the employed circuit nodes do not demand any endorsement and pseudonym signatures. 
Traffic sent to these whitelisted destinations through the circuit remains anonymous in the current AC networks sense.

\pparagraph{Non-listed destinations:}
For destinations that are not listed in the exit-node whitelist directory, the user has to use \sysname while building the circuit to it;
otherwise, the exit node will drop her requests to the non-listed destinations.

We emphasize that \sysname is {\em not} an ``all-or-nothing'' design alternative: it allows an AC network to conveniently disable the complete verification and logging mechanisms for some pre-selected destinations. In particular, an exit node with ``Sorry, it is an anonymity network, no logs'' opinion can still whitelist the whole Internet, while others employ \sysname for non-whitelisted sites. The use of BackRef is transparent, and users can choose if they wish to use a BackRef node for their circuits.


\begin{figure}[!h]
\centering
\framebox[\columnwidth]{\begin{minipage}[!h]{0.95\columnwidth} \small{
\upon an input (\setup):
\begin{algorithmic}
\STATE Generate an asymmetric key pair $(\sk,\pk) \leftarrow G$.
\STATE send a cell $(\register, N,\pk)$ to the \FREG functionality
\STATE {\bf wait for} a cell $(\registered,  \langle N_j, \pk_j \rangle_{j=1}^n)$ from \FREG
\STATE output $(\ready, {\cal N} = \langle N_j \rangle_{j=1}^n)$
\end{algorithmic}

\upon an input $(\createcircuit,\Parties = \langle N, \langle N_j \rangle_{j=1}^\ell\rangle)$:\hfill

\begin{algorithmic}
\STATE store $\Parties$ and $\Circuit \leftarrow \langle N\rangle$; 
call $\ExtendCircuit(\Parties,\Circuit)$
\end{algorithmic}

\upon an input $(\send, \Circuit = \langle N\link{\cid_1}N_1\link{~}\cdots N_{\ell}\rangle, m)$:\hfill

\begin{algorithmic}

\STATE look up the keys $(\langle k_j \rangle_{j=1}^{\ell})$ for $\cid_1$
\STATE $O \leftarrow \WrapOnion(m, \underline{\sigma_{\pseudonym_\ell}, \ts}, (k_j)_{j=1}^{\ell})$; $\mathit{Used}(\cid_1)$++
\STATE send a cell $(\cid_1,\relay, O)$ to $N_{1}$ over \FSCS

\end{algorithmic}

\upon {receiving a cell $(\cid, \create, \pseudonym, \underline{\sigma_i, \ts})$ from $N_i$ over \FSCS:}\hfill

\begin{algorithmic}
\IF{ \underline{$\mathit{Verify(\sigma_i, \pk_{N_i})}$}}
\STATE $\langle Y, k_{\new} \rangle \leftarrow \Respond(\pk_N,\sk_N, \pseudonym)$
\STATE store $\Circuit \leftarrow \langle N_i\link{\cid, k_{\new}}N\rangle$
\STATE \underline{store $Log \leftarrow \langle H(\pseudonym),IP_{N_i}\pseudonym, \sigma_i, \ts \rangle$}
\STATE send a cell $(\cid,\created,  Y,t)$ to $N_{i}$ over \FSCS
\ENDIF
\end{algorithmic}

\upon {receiving a cell $(\cid, \created, Y, t)$  from $N_i$ over \FSCS:}\hfill

\begin{algorithmic}
\IF {$\prev(\cid) = (N',\cid',k')$} 
\STATE $O\leftarrow \WrapOnion(\langle \extended, Y, t \rangle, k')$
\STATE send a cell $(\cid',\relay,O)$  to $N'$ over \FSCS
\ELSIF {$\prev(\cid) = \bot$}  
\STATE $k_\new \leftarrow \ComputeKey(\pk_i, Y, t)$
\STATE update $\Circuit$ with $k_\new$; call $\ExtendCircuit(\Parties,\Circuit)$
\ENDIF
\end{algorithmic}

\upon {receiving a cell $(\cid, \relay,O)$  from $N_i$ over \FSCS:}\hfill

\begin{algorithmic}
\IF  {$\prev(\cid) = \bot$}
\IF {$\getkey(\cid) = (k_j)_{j=1}^{\ell'}$}  
\STATE $(\type, m)$ {\bf or} $O \leftarrow \UnwrapOnion(O, (k_j)_{j=1}^{\ell'})$ 
\STATE $(N',\cid') $ {\bf or} $\bot\leftarrow \nxt(\cid)$ 
\ENDIF
\ELSIF {$\prev(\cid) = (N',\cid', k')$}  
\STATE $O \leftarrow \WrapOnion(O, k')$ /* a backward onion */
\ENDIF
\STATE {\bf switch} (\type)
\STATE{\bf case} {\extend}: 
\STATE \hspace{2ex}get $\langle N_{\nxt},  \pseudonym, \underline{ \sigma_{\pseudonym_i},\ts}\rangle $ from $m$; $\cid_{\nxt} \ot \{0,1\}^\kappa$
\IF{\underline{$\mathit{Verify(\sigma_{\pseudonym_i}, \pseudonym_i)}$}}   
\STATE \hspace{2ex}update $\Circuit \leftarrow\langle N_i \link{\cid, k} N\link{\cid_\nxt}N_\nxt\rangle$
\STATE \hspace{2ex}\underline{store $Log \leftarrow \langle H(\pseudonym), IP_{N_i} \pseudonym, \sigma_{\pseudonym_i}, \ts \rangle$}
\STATE \hspace{2ex}send a cell $(\cid_{\nxt},\create, \pseudonym)$ to $N_{\nxt}$ over \FSCS
\ENDIF
\STATE {\bf case} {\extended}: 
\STATE \hspace{2ex}get $\langle Y, t\rangle$ from $m$; get $N_{\added}$ from $(\Circuit, \Parties)$
\STATE \hspace{2ex}$k_{\added} \leftarrow \ComputeKey(\pk_{\added},  Y, t)$
\STATE \hspace{2ex}update $\Circuit$ with $( k_{\added})$;
call $\ExtendCircuit(\Parties,\Circuit)$
\STATE {\bf case} {\data}:
\STATE \hspace{2ex}{\bf if} {(N = OP)} {\bf then} output $(\received,\Circuit, m)$
\STATE \hspace{2ex}{\bf else} {\bf if} $m = (S, m', \underline{\sigma_{\pseudonym},\ts})$
\STATE \hspace{2ex}  \underline { store $Log  \leftarrow  \langle H(m), IP_{N_i}, \pseudonym, \sigma_\pseudonym, \ts \rangle$}
\STATE \hspace{4ex} generate or lookup the unique $\sid$ for $\cid$
\STATE \hspace{4ex} send $(N, S, \sid, m')$ to the network
\STATE {\bf case} {\default}: /*encrypted forward/backward onion*/
\STATE \hspace{2ex}send a cell $(\cid', \relay,O)$ to $N'$ over \FSCS
\end{algorithmic}

\upon {receiving a msg $(\sid, m)$ from $\FNET{q}$:}\hfill

\begin{algorithmic}
\STATE  get $\Circuit  \leftarrow  \langle N' \link{\cid, k} N\rangle$ for $\sid$; $O \leftarrow \BConstructOnion( m,k)$
\STATE send a cell $(\cid,\relay, O)$ to $N'$ over \FSCS
\end{algorithmic}

%
%

}
\end{minipage}
}
\vskip 0.5em 
\scriptsize{Without circuit destruction.}
\caption{\label{figure:or}$\POR$ with \sysname for Party $N$}
\vskip -6.5em 
\end{figure}

%
%


\begin{figure}[th!]
\centering
\framebox[\columnwidth]{\begin{minipage}[!h]{0.98\columnwidth} \small{
{\noindent$\ExtendCircuit(\Parties =  \langle N_j\rangle_{j=1}^\pathlength,
\Circuit = \langle N\link{\cid_1,k_1}N_1\link{k_2}\cdots N_{\ell'}\rangle)$:}\hfill

\begin{algorithmic}
\STATE determine the next node $N_{\ell' +1}$ from \Parties and \Circuit
\IF {$N_{\ell' +1} = \bot$}
\STATE output $(\created, \langle N\link{\cid_1}N_1\link{~}\cdots N_{\ell'}\rangle)$
\ELSE
\STATE $\pseudonym \leftarrow \Initiate(\pk_{N_{\ell' +1}} , N_{\ell' +1})$
\IF {$N_{\ell' +1} = N_{1}$}
\STATE $\cid_1 \ot \{0,1\}^\kappa$
\STATE send a cell $(\cid_1,\create, \pseudonym)$ to $N_{1}$ over \FSCS
\ELSE
\STATE $O \leftarrow \WrapOnion(\{\extend,N_{\ell' +1}, \pseudonym, \underline{\sigma_{\pseudonym_{\ell'}},\ts}\}, (k_j)_{j=1}^{\ell'})$
\STATE send a cell $(\cid_{1}, \relay, O)$ to $N_{1}$ over \FSCS
\ENDIF
\ENDIF
\end{algorithmic}
}
\end{minipage}
}
\caption{\label{figure:SubProt2}Subroutine for \POR with \sysname for $N$}
\end{figure}

\begin{figure}[th!]
\centering
\framebox[\columnwidth]{\begin{minipage}[!h]{0.95\columnwidth} \small{

\upon { a verification request $(m)$:}\hfill

\begin{algorithmic}
\IF {$LookupLog(H(m)) = \bot$}
\STATE $TraceFail(m)$
\ELSE
\STATE { get $Log  \leftarrow  \langle H(m), N_\prev, \pseudonym, \sigma, \ts \rangle $ for $H(m)$}

\IF {((N = N$_1$) \& $Verify(\sigma, \pseudonym)$)}
\STATE output $(\pseudonym, N_\prev)$ 
\ELSE 
\STATE get $Log  \leftarrow  \langle H(\pseudonym), N_{N_{\prev}}, pk_{N_{\prev}}, \sigma^\prime, \ts \rangle $ for $H(X)$

\IF {($Verify(\sigma, \pseudonym)$ \& $Verify(\sigma^\prime, pk_{N_{\prev}})$)}
\STATE output $(\pseudonym, N_{\prev})$
\ELSE 
\STATE $TraceFail(m)$
\ENDIF
\ENDIF
\ENDIF

\end{algorithmic}
}
\end{minipage}
}
\caption{\label{figure:SubProt}Backward Traceability Verification}
\end{figure}

\subsection{Pseudocode}

In this subsection, we present pseudocode for the OR protocol with \sysname 
extending the OR pseudocode developed by Backes \etal~\cite{UC-OR}
following the Tor specification~\cite{tor-spec}.
We highlight our changes to their original (\POR) protocol pseudocode from \cite{UC-OR} 
by underlining those.
Our pseudocode formalism 
demonstrates that our modification the original OR protocol are minimal.
It also forms the basis for our applied pi calculus \cite{OEquivalence} based OR model in Section~\ref{sec:Analysis}. \vspace{3ex}
In the pseudocode, an OR node maintains a state for every protocol execution
and responds (changes the state and/or sends a message) 
upon receiving a message. 
There are two types of messages that the protocol employs: 
the first type contains {\em input} and   {\em output} actions, 
which carry respectively the user inputs to the protocol, 
and the protocol outputs to the user;
the second message type is a network message (a cell in the OR literature), 
which is to be delivered by one protocol node to another.

In onion routing, a directory server maintains the list of valid OR nodes and the respective public keys.
A functionality $\FREG$ abstracts this directory server.
Each OR node initially computes its
long-term keys $(\sk, \pk)$ (for both 1W-AKE and signature schemes) and registers the
public part at $\FREG$. 

For ease of exposition, cryptographically important Tor cells are considered in the protocol.
This includes \create, \created and \destroy
cells among control cells, and \data, \extend and \extended cells among \relay cells.
There are two input messages \createcircuit and \send, 
where the user uses \createcircuit to create OR circuits and 
uses \send to send  messages $m$ over already-created circuits.



\begin{sloppypar}
The \ExtendCircuit function defined in Figure~\ref{figure:SubProt2}
presents the circuit construction description from Section~\ref{sec:OROverview} in a pseudocode form.
Circuit IDs $(\cid \in \{0,1\}^\kappa)$ associate two consecutive circuit nodes in a circuit. 
The terminology 
$\Circuit = N_{i-1}\link{\cid_i, k_i}N_i\link{\cid_{i+1}}N_{i+1}$,
says that $N_{i-1}$ and $N_{i+1}$ are respectively the predecessor and successor of $N_i$ in a circuit \Circuit.
$k_i$ is a session key between $N_i$ and the OP, while 
the absence of $k_{i+1}$ indicates that a session key between
$N_{i+1}$ and the OP is not known to $N_i$; analogously the absence of
a circuit id $\cid$ in that notation means that only the first circuit
id is known, as for OP, for example.
Functions \prev and \nxt on \cid correspondingly return information about the 
predecessor or successor of the current node with respect to \cid; 
e.g., $\nxt(\cid_i)$ returns $(N_{i+1},\cid_{i+1})$
and $\nxt(\cid_{i+1})$ returns $\bot$.
The OP passes on to the user $\langle N\link{\cid_1}N_1\link{~}\cdots N_{\ell}\rangle$.
\end{sloppypar}

Within a circuit, a user's OP (onion proxy) and the exit node use \relay cells created using wrapping algorithm 
\WrapOnion to tunnel end-to-end commands and communication.
The exit nodes use the streams to synchronize communication between the network and a circuit $\Circuit$.
It is represented as \sid in the pseudocode.
End-to-end communication between OP and the exit node
happens with a \WrapOnion call with multiple session keys and a series of \UnwrapOnion calls with individual session keys.
Cells are exchanged between OR nodes over a secure and authenticated channels,
e.g., a TLS connection,
and they are modeled a secure channel functionality \FSCS~\cite{Can01}.
%
Circuit destruction remains exactly the same in our case, 
and we {\em omit} it in our pseudocode and refer the readers to \cite{UC-OR} details. 

In Figure~\ref{figure:SubProt}, we formalize the backward traceability verification of \sysname. 
Here, function $LookupLog$ determines an entry from the log index by its input. 
Function $Verify$ performs signature verification, 
while function $TraceFail$ outputs that a valid log entry does not exists at node $N$.

\section{Systems Aspects and Discussion}\label{sec:TSysAspects}

\paragraph{Communication overhead}
Communication overhead for \sysname is minimal: every circuit creation, circuit extension, and stream request carries a $32$ byte BLS signature and additional $4$ byte timestamp.

\paragraph{Computation overhead}
In a system with \sysname, every node has to verify a signature and generate another.
Using the pairing-based cryptography (PBC) library, a BLS signature generation takes less than 1ms
while a verification asks for nearly 3ms for 128-bit security on a commodity PC
with an Intel i5 quad-core processor with 3.3 GHz and 8 GB RAM.
Signing and verification time (and correspondingly system load) 
can be further reduced using faster signature schemes (\eg~\cite{fastSignature}).

\paragraph{Log storage} \sysname requires nodes to maintain logs of cryptographic information for potential use by law enforcement. These logs are not innocuous, and the implications of publicly disclosing a record need to be considered. The specificity of the logs should be carefully designed to balance minimal disclosure of side-information (such as specific timings) while allowing flows to be uniquely identified. It must also be possible to reconstruct the logged data from the types of information available to law enforcement. The simplest entry would contain the destination IP, source (exit node) IP, a coarse timestamp, as well as the signature. Logs should be maintained for a pre-defined period and then erased.

No single party can hold the logs without entrusting the anonymity of all users to this entity. The OR nodes can retain the logs themselves. This, however, would require law enforcement to acquire the logs from every such node and consequently involve the nodes in the investigation---a scenario that may not be desirable. Furthermore, traceability exposes nodes of all types, not just exit nodes, to investigation. We are aware of a number of entities who deliberately run middle nodes in Tor to avoid this exposure. An alternative is to publish encrypted logs, where a distributed set of trustees share a decryption key and act as a liaison to law enforcement, while holding each other accountable by refusing to decrypt logs of users who have not violated the traceability policy. Such an entity acts in a similar fashion to the group manager schemes based on group signatures~\cite{SelectTraceable}.
%

\paragraph{Non-cooperating nodes} Given the geographic diversity of the AC networks, it is always possible that some proxy nodes cooperate with
the \sysname mechanism, while others do not. The repudiation property of \sysname ensures that a cooperating node can always at least
correctly shift liability to a non-cooperating node.
Furthermore, such a cooperating node may also {\em reactively} decide to block
any future communication from the non-cooperating node as a policy.

\paragraph{ISP as a trusted party}
In the absence of a PKI for users, to solve the last mile problem, our protocol has to rely on some trust mechanism to prove the linkability between the IP address of the user and the entry node pseudonym. For this purpose, we consider an ISP with packet attestation mechanism ~\cite{packetsISP} to be a proper solution that adds a small overhead for the existing ISP infrastructure and at the same time does not harm any of the properties provided by the anonymity network. In some countries there is an obligation for the ISPs to retain data that identify the user, in others where the ISPs are not obligated by law, it is a common practice. 
The protocol is designed in a way that the ISP has to attest only to the {\em ClientKeyExchange} message (this message is a part of the TLS establishing procedure, and also is public and not encrypted message) which is used to establish the initial TLS communication. This message does not reveal any sensitive information related with the identity of the user. By its design, we reuse this message as a pseudonym for the entry OR node. 


\begin{figure*}[t!]
\centering
\includegraphics[trim=4.4cm 0cm 4.1cm 0cm,clip=true,angle=-90,scale=0.58]{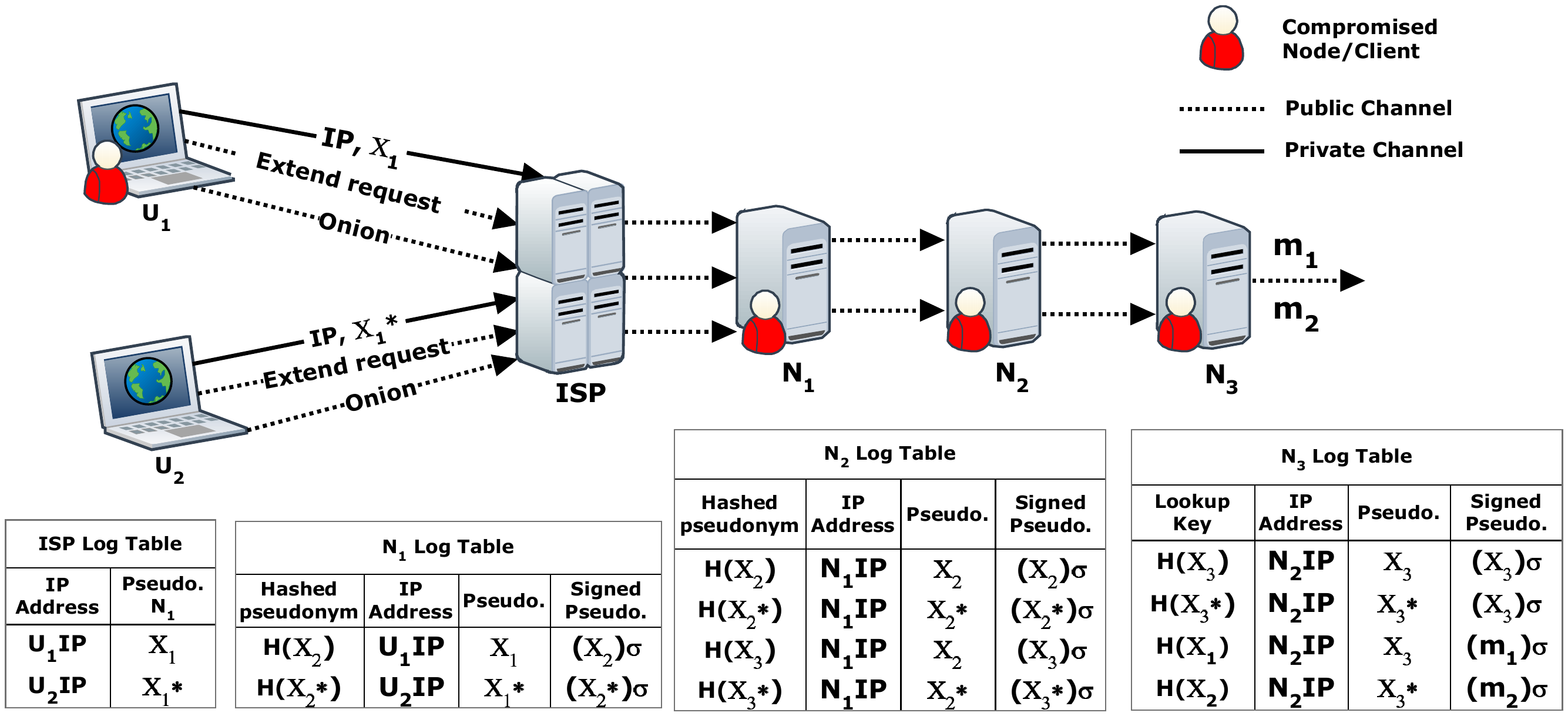}
\caption{No False Accusation adversarial model}
\label{NoFalseAcc}
\end{figure*}

\section{Security Analysis}
\label{sec:Analysis}
In this section we present a formal security analysis of \sysname. 
We model our protocol from the previous section (in a restricted form)
in the applied pi calculus \cite{OEquivalence} and 
verify the important properties anonymity, backward traceability, no forward traceability, and no false accusation 
with ProVerif~\cite{Proverif}, a state-of-the-art automated theorem prover
that provides security guarantees for an unbounded number of protocol sessions.
We model backward traceability and no false accusation as trace properties, and
anonymity and no forward traceability as observational equivalence relations.
The ProVerif scripts used in the analyses are publicly available~\cite{Proverif:scripts}.

\paragraph{Basic Model}
We model the OR protocol in the applied pi calculus to use circuits of length three (\ie one user and three nodes); 
the extension to additional nodes is straightforward.
To prove different security properties we upgrade the model to use additional processes and events.
The event contents used to decorate the various steps in the OR protocol as well as \sysname mechanism
follow the pseudocode from the previous section. 
We also involve an ISP between the user and the entry node, which participate in the protocol as a trusted party. The ISP is honest and can prove the existence of a communication channel between the user and the entry node. This channel is modeled to be private, preventing any ISP log forgeries. 
The cryptographic log collection model is designed in a decentralized way such that nodes retain the logs themselves in a table that is inaccessible to the adversary.

We model the flow of the pseudonyms and the onion, together with the corresponding verification.
However, we do not model the underlying, cryptographically verified 1W-AKE \ntor protocol, 
and assume that the session key between the user and the selected OR process is exchanged securely. 
The attacker is a standard Dolev-Yao active adversary with full control over the public channels:
It learns everything ever send on the network, and can create and insert messages on the public channels.
It also controls network scheduling.

\paragraph{Backward Traceability}
The essential goal of our protocol is to trace the source of the communication stream starting from an exit node. We verify that the property of backward traceability arrives from the correctness of the (backward) traceability verification mechanism. 

%
%
%

\begin{sloppypar}
The correctness property can be formalized in ProVerif notation as follows:


\begin{equation}\label{Corectness}
  \begin{split}
TraceUser(IP) \Longrightarrow (LookupISP(X_1, IP) \Longrightarrow \\ 
( RevealPredecessorU(IP)) \Longrightarrow \\
(RevealPredecessor(ipN1) ) \Longrightarrow \\ 
(RevealPredecessor(ipN2)) \wedge CheckSignature \\ 
\wedge  LookupN3(m)))
  \end{split}
\end{equation}

where the notation A $\Longrightarrow$ B denotes the requirement that the event A must be preceded by a event B. In our protocol, the property says that the user is traced if and only if all nodes in the circuit verifiably trace their predecessors. 
The traceability protocol $P$ starts with the event $LookupN3(m)$ which means that for a given message $m$ (stream request) the verifier consults the log, and if such request exists, it checks the signature $CheckSignature$. Finally when all these conditions are fulfilled, the verifier reveals the identity of the predecessor node $RevialPredecessor(ipN2)$ (\ie the middle node). This completes the nested correspondence $ (CheckSignature \wedge LookupN3(m) \wedge  RevealPredecessor(ipN2))$ which verifiably traces $N_2$.
In a similar fashion, after all conditions are fulfilled, the verifier traces $N_1$ and the user $U$.
\end{sloppypar}

\begin{sloppypar}
After the identity of $U$ is revealed, the verifier lookup into the evidence table of the ISP ($LookupISP$) to prove the connection between the identity of the user $IP$ and the pseudonym of the entry node $X_1$. If such record exist into the table, the address of the user is revealed and the event $TraceUser(IP)$ is executed.

\end{sloppypar}

\textit{Theorem:} The trace property defined in equation~(\ref{Corectness}) holds true for all possible executions of process P.

\begin{proof}
Automatically proven by ProVerif.
\end{proof}

\begin{figure*}[t!]
\centering
\includegraphics[trim=5.0cm 0cm 3.3cm 0cm,clip=true,angle=-90,scale=0.53]{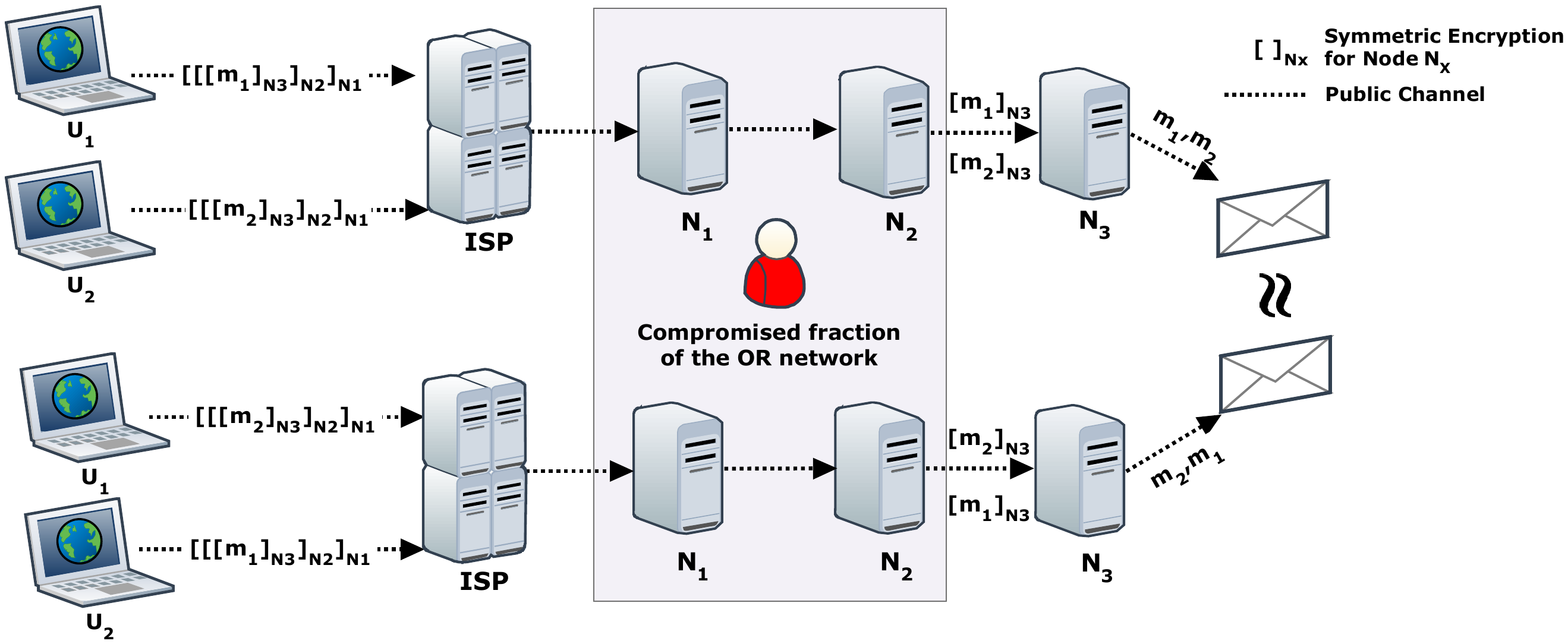}
\caption{Anonymity Game}
\label{Anonymity} 
\end{figure*} 

\paragraph{No false accusation}
There are two aspects associated with false accusations:

\begin{enumerate}
\item It should not be possible for a malicious node $N_A$ to trace a communication stream
 to an OR node $N_C$ other than to its predecessor in the corresponding circuit. 
 Informally, to break this property, $N_A$ has to be obtain a signature of $N_C$
 on a particular pseudonym associated with the circuit. This requires $N_A$ to
 forge a signature for $N_C$, which is not possible due to the unforgeability property of the signature scheme.

\item It should not be possible for a malicious node $N_A$ to trace a communication stream
to a circuit $C_1$ other than the circuit $C_2$ employed for the communication stream.
Consider a scenario where two concurrent circuits ($C_1$ and $C_2$), 
established by two different users $U_1$ and $U_2$, pass through a malicious node $N_A$. 
Suppose that $N_A$ collaborates with $U_2$ who is misbehaving and have used the OR network for a criminal activities. 
To help $U_2$ by falsely accusing a different predecessor, $N_A$ must forge two signatures: 
To link two pseudonyms ${\pseudonym_1}_{i-1}$ and ${\pseudonym_2}_{i}$ from circuits $C_1$ and $C_2$ respectively, 
$N_A$ has to forge the pseudonym signature on ${\pseudonym_2}_{i}$ with ${\pseudonym_1}_{i-1}$ as a public key,
 or he has to know the temporal signing key pair for the predecessor in $C_1$. 
\end{enumerate}

Intuitively, the first case is ruled out by the unforgeability property of the signature scheme.
We model the later case as a trace property.
Here, even when $N_A$ collaborates with $U_2$, 
it cannot forge the signed pseudonym received from its predecessor.
The property remains intact as long as one of nodes on $C_1$ and the packet attesting ISP~\cite{packetsISP}
remains uncompromised. In absence of a PKI or credential system for users, the last condition is unavoidable.

%

We formalize and verify the latter case of the property in an adversarial model where the attacker has compromised one user ($U_1$ or $U_2$). Figure~\ref{NoFalseAcc} provide a graphical representation of the protocol $P$. We upgrade the basic model involving additional user $U_2$ who sends additional message $m_2$. As mentioned before, to simulate the packet attesting mechanism~\cite{packetsISP} we involve a honest ISP between the user and the entry node. The ISP only collects data that identifies the user (IP address of the user) and the pseudonym for the entry node ($\pseudonym_1$) which is send in plain-text. The adversary does not have an access to the log stored by the ISP i.e. cannot read or write anything into the log table. We want to verify that for all protocol execution the request $m_i$ cannot be associated with any user $U_i$ other than the originator.

To formalize the no false accusation property in ProVerif, we model security-related protocol events with logical predicates. The events $CorrN1$, $CorrN2$, $CorrN3$ in the protocol occur only when the OR nodes $N_1$, $N_2$, $N_3$, respectively, are corrupted. The event CorrISP defines the point of the protocol where the ISP is corrupted.
The no false accusation property is formalize as the following policy: 

%


\begin{equation}\label{NoFalseAccusation}
  \begin{split}
    Accuse(IP, m) \Longrightarrow \\
 (CorrN1 \wedge CorrN2 \wedge CorrN3 \wedge CorrISP).
  \end{split}
\end{equation}


This policy says that if a user with address IP is falsely accused for a message $m$ i.e. $Accuse(IP,m)$, then indeed all of the parties in the protocol has to be corrupted.

%
%

\textit{Theorem:} The trace property defined in equation (\ref{NoFalseAccusation}) holds true for all possible executions of process P.

\begin{proof}
Automatically proven by ProVerif.
\end{proof}




\begin{figure*}[t!]
\centering
\includegraphics[trim=8.5cm 1.4cm 7.8cm 1.4cm,clip=true,angle=-90,scale=0.67]{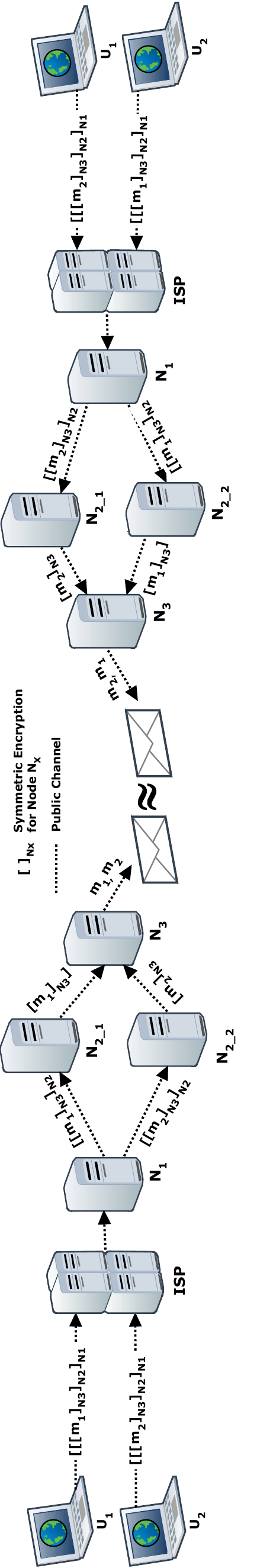}
\caption{No Forward Traceability}
\label{NoForwardTrace}
\end{figure*}

\hyphenation{ano-nym-ity}
\paragraph{Anonymity}
%
We model this property as an observational equivalence relation between two processes that are replicated an unbounded numbers of time and execute in parallel.  In the first process $P$, users $U_1$ and $U_2$ send two messages $m_1$ and $m_2$, respectively. While in the second process Q the two messages are swapped. If the two defined processes are observationally equivalent (P $\approx$  Q), then we say that the attacker cannot distinguish between $m_1$ and $m_2$ i.e. cannot learn which message is sent by which user. In our scenario we assume that the attacker can compromise some fraction of the OR node, but not all. Figure~\ref{Anonymity} provide a graphical representation of the anonymity game where the exit node $N_3$ is honest.  
The game works as follows:
\begin{enumerate}

\item $U_1$ and $U_2$ create an onion data structure $O_1$ and $O_2$, respectively, intended for $N_3$ and send via previously built circuits  $C_1$ ($U_1\leftrightarrow N_1\leftrightarrow N_2\leftrightarrow N_3$) and  $C_2$($U_2\leftrightarrow N_1\leftrightarrow N_2\leftrightarrow N_3$).  Nodes communicate between each other through public channel. 

\item Two of the intermediate nodes are corrupted and the attacker has full control over them. The intermediate compromised nodes (in our case $N_1$ and $N_2$) remove one layer of encryption from $O_1$ and $O_2$ and send the onion to the exit node $N_3$.
\item After receiving these two onions from the users $U_1$ and $U_2$ and possibly other onions from compromised users, the exit OR node $N_3$ remove the last layer of the encryption and publish the message on a public channel.
\end{enumerate}

Note that the ISP does not affect the anonymity game and only act as a proxy between the users and the outside world.
For the anonymity verification, we assume that user $U_1$ and user $U_2$ are honest and they follow the protocol. Nevertheless, the action of any compromised user and honest users can be interleaved in any order.

\textit{Theorem: }The observational equivalence relation $P \approx Q$ holds true.

\begin{proof}
Automatically proven by ProVerif.
\end{proof}


Notice that the evidence records here inherently break anonymity: anybody with access to logs of the entry, middle, and exit nodes of a circuit can break the user anonymity. Therefore, traceability logs have to be indexed and individually encrypted using an appropriate trust-enforcing mechanism. In Section~\ref{sec:TSysAspects}, we discuss the possible solutions.

\paragraph{No forward traceability}
The evidence log of the backward traceability protocol in \sysname does not store any information (\ie IP addresses) that can identify or verifiably reveal the identity of a node's successor. The log contains only the pseudonym for the successor node which does not reveal anything about the identity of the node.

%
%
%
%
%

We formalize this property as an observational equivalence relation between two distinct processes and verify that an adversary cannot distinguish them. Figure~\ref{NoForwardTrace} provides a graphical representation of the game. To prove the observational equivalence, we model a scenario with concurrent circuit executions. 
In this game, the adversary can corrupt parties and extract their secrets only after the message transmission over the circuit has completed.
For this game, our model involves an additional middle node and user $ U_2$. Two users $U_1$ and $U_2$ send two different messages $m_1$ and $m_2$ via two circuits. 
We verify that it is impossible for an attacker to deduce any meaningful information about the successor node for a particular request.
Our game works as follows:
\begin{enumerate}
\item $U_1$ and $U_2$ start the protocol and constructs two different circuits $C_1$($U\leftrightarrow N_1\leftrightarrow N_2\leftrightarrow N_3$) and $C_2$($U\leftrightarrow N_1\leftrightarrow N_2^*\leftrightarrow N_3$), respectively with adequate values $(x_1, x_2, x_3)$ for a circuit $C_1$ and $(x_1^\prime, x_2^\prime, x_3^\prime)$ for $C_2$.

\item $U_1$ and $U_2$ create an onion data structure $O_1$ and $O_2$ and send to the exit node $N_3$ via previously built circuits $C_1$ and $C_2$. Nodes communicate between each other through public channels.
 
\item After receiving the two onions from the users and possibly other onions from compromised users, $N_3$ removes the last layer of the encryption and publishes the messages on a public channel.

\item After protocol completion, the entry node $N_1$ is compromised and the adversary obtains the evidence log.

\end{enumerate}

In the first process $P$, $U_1$ sends $m_1$ and $U_2$ sends $m_2$, while the process is reversed process Q. For the no forward traceability verification, we assume that all other parties in the protocol remain honest, except the compromised $N_1$. For example, if two neighbor nodes are compromised, the no forward traceability can be easily broken with activating the backward traceability mechanism.

\textit{Theorem: }The observational equivalence relation $P \approx Q$ holds true.

\begin{proof}
Automatically proven by ProVerif.
\end{proof}

Finally, to the best of our knowledge, 
our formal analysis is the first ProVerif-based analysis of the OR protocol;
it can be of independent interest towards formalizing and verifying other properties of the OR protocol.

\section{Conclusions}\label{sec:Conclusions}
In this paper, we presented \sysname, an accountability mechanism for AC networks that provides practical repudiation for the proxy nodes, allowing selected outbound traffic flows to be traced back to the predecessor node. It also provides a full traceability option when all intermediate nodes are cooperating. While traceability mechanisms have been proposed in the past, \sysname is the first that is both compatible with low-latency, interactive applications (such as anonymous web browsing) and does not introduce new trusted entities (like group managers or credential issuers). \sysname is provably secure, requires little overhead, and can be adapted to a wide range of anonymity systems. 

\bibliographystyle{IEEEtran}
\bibliography{bib/tor}

\appendices
\section{Bilinear Pairings}\label{App:BLS}
In this section, we briefly review bilinear pairings. For more detail see \cite{ECC} and references therein.

Consider two additive cyclic groups $\g_1$ and $\g_2$ and a multiplicative cyclic group $\g_T$, all of the same prime order $p$.  A
bilinear map $\e$ is a map $\e : \g_1 \times \g_2 \to \g_T$ with the 
following properties.

\begin{sloppypar}
\begin{LaTeXdescription}
\item [Bilinearity:] For all $P \in \g_1$, $Q \in \g_2$ and $a,b
  \in \mathbb{Z}_p$, $\e(P^a,Q^b) = \e(P,Q)^{ab}$.
  
\item [Non-degeneracy:] The map does not send all pairs in $\g_1
  \times \g_2$ to unity in $\g_T$.    
  
\item [Computability:] There is an efficient algorithm to compute
  $\e(P,Q)$ for any $P \in \g_1$ and $Q \in \g_2$.

\end{LaTeXdescription}
\end{sloppypar}

\section{BLS Signatures}\label{App:BLS}
In this section, we briefly review BLS signatures. For more detail see \cite{BLSsig} and references therein.

Consider two Gap co-Diffie-Hellman groups (or co-GDH group) $\g_1$ and $\g_2$ 
and a multiplicative cyclic group $\g_T$, all of the same prime order $p$, 
associated by a bilinear map~\cite{ECC} $\e : \g_1 \times \g_2 \to \g_T$.
Let $g_1$, $g_2$, and $g_T$ be generators for $\g_1$, $\g_2$, and $\g_T$ respectively
and let a full-domain hash function $H : \left\{0,1\right\}^* \to \g_1$. 
The BLS signature scheme~\cite{BLSsig} comprises three algorithms, \textit{Key Generation}, 
\textit{Signing} and \textit{Verification} defined as follows:

\begin{LaTeXdescription}
\item [Key Generation:] Choose random $\sk \in_R\mathbb{Z}_p  $ and compute $\pk = g_2^\sk$. The private key is $\sk$, and the public key is $\pk$.
\item [Signing:] Given a private key $\pk \in \mathbb{Z}_p$, and a message $m \in \{0,1\}^*$, compute $h = H(m) \in \g_1$ and signature $\sigma = h^\sk$, where $\sigma \in \g_1$. 
\item [Verification:] Given a public key $\pk \in \g_2$, message $m \in \{0,1\}^*$, and signature $\sigma \in \g_1$, compute $h = H(m) \in \g_1$ and verify that $(g_2, \pk, h, \sigma)$ is a valid co-Diffie-Hellman tuple. 
\end{LaTeXdescription}

\section{1W-AKE Protocol}\label{sec:ntor}
Until recently, Tor has been using an authenticated Diffie-Hellman (DH) key agreement protocol called the Tor authentication protocol (TAP),
where users' authentication challenges are encrypted with RSA public keys of OR nodes. However, this atypical use of RSA encryption is found to be inefficient in practice, and several different interactive and non-interactive (one-way authenticated) key agreement (1W-AKE) protocols
have been proposed in the literature~\cite{LaPa_07, ntor, Certificateless, Kate07pairing-basedonionimproved, KateSphinx, Ace}.
TAP has recently been replaced by the \textit{ntor} protocol by Goldberg, Stebila and Ustaoglu~\cite{ntor}. The \textit{ntor} protocol is in turn derived from a protocol by {\O}verlier and Syverson~\cite{LaPa_07}.


The protocol \ntor~\cite{ntor} is a 1W-AKE protocol between two parties $P$ (client) and $Q$ (server), 
where client $P$ authenticates server $Q$. 
Let $(\pk_Q, \sk_Q)$ be the static key pair for $Q$. We assume that $P$ holds $Q$'s certificate $(Q,\pk_Q)$. $P$ initiates 
an \ntor session by calling the $\Initiate$ function and sending the output message $m_P$ to $Q$. 
Upon receiving a message $m'_P$, server $Q$ calls the \Respond function and sends the output message $m_Q$ to $P$.
Party $P$ then calls the \ComputeKey function with parameters from the
received message $m'_Q$, and completes the \ntor protocol. We assume a
unique mapping between the session ids $\Psi_P$ of the \cid in \POR.

\begin{figure}[t]
\centering
\framebox[\columnwidth]{\begin{minipage}[!h]{0.98\columnwidth} \small
\noindent $\Initiate(\pk_Q, Q)$:\smallskip
  \begin{compactenum}
  \item Generate an ephemeral key pair $(x, X \leftarrow g^x)$.
  \item Set session id $\Psi_P \leftarrow H_{\state}(X)$.
  \item Update $\state(\Psi_P) \leftarrow (\ntor, Q, x, X)$.
  \item Set $m_P \leftarrow (\ntor, Q, X)$.
  \item Output $m_P$
.
  \end{compactenum}
\medskip
\noindent $\Respond(\pk_Q, \sk_Q, X)$:\smallskip
  \begin{compactenum}
 \item Verify that $X \in G^*$.
  \item Generate an ephemeral key pair $(y, Y \leftarrow  g^y)$.
  \item Set session id $\Psi_Q \leftarrow H_\state(Y)$.
  \item Compute $(k', k) \leftarrow H({X}^y, X^{\sk_Q}, Q, X, Y,\ntor)$.
  \item Compute $t_Q \leftarrow H_{mac}(k', Q, Y, X, \ntor,\server)$.
  \item Set $m_Q  \leftarrow (\ntor, Y, t_Q)$.
  \item Set $\mathit{out} \leftarrow (k, \star, X, Y, \pk_Q)$, where $\star$ is the anonymous party symbol.
  \item Delete $y$ and output $m_Q$.
  \end{compactenum}
\medskip
\noindent $\ComputeKey(\pk_Q, \Psi_P, t_Q, Y)$:\smallskip
  \begin{compactenum}
  \item Retrieve $Q$, $x$, $X$ from $\state(\Psi_P)$ if it exists.
  \item Verify that $Y \in G^*$.
  \item Compute $(k', k) \leftarrow H(Y^x, pk_Q^x, Q, X, Y,\ntor)$.
  \item Verify $t_Q = H_{mac}(k', Q, Y, X, \ntor,\server)$.
  \item Delete $\state(\Psi_P)$ and output $k$. 
  \end{compactenum}
If any verification fails, the party erases all session-specific
information and aborts the session.
\end{minipage}
}
\caption{The \ntor protocol}\label{figure:ntor}
\vskip -1em
\end{figure}

%


\end{document}